\documentclass[twocolumn,showpacs,preprintnumbers,amsmath,amssymb,floatfix]{revtex4}
\usepackage{graphicx}
\usepackage{dcolumn}
\usepackage[latin1]{inputenc}
\usepackage{color}

\begin{document}


\title{Beyond Bloch: A Theoretical Blueprint for Conjugated Polymer
Optoelectronics} 

\author{Miguel Lagos}
\email{mlagos7@gmail.com}
\affiliation{Departamento de F{\'\i}sica, Facultad de Ciencias,
Universidad de Chile, Casilla 653, Santiago, Chile.}

\author{Miguel Kiwi}
\email{m.kiwi.t@gmail.com}
\affiliation{Departamento de F{\'\i}sica, Facultad de Ciencias,
Universidad de Chile, Casilla 653, Santiago, Chile 7800024, and
Centro para el Desarrollo de la Nanociencia y la Nanotecnolog{\'i}a,
CEDENNA, Avenida Ecuador 3493, Santiago, Chile 9170124.}

\author{Rodrigo Paredes}
\email{raparede@utalca.cl}
\affiliation{Facultad de Ingenier\'\i a, Universidad Finis Terrae,
Avenida Pedro de Valdivia 1509, Providencia, Santiago, Chile.}

\date{today}

\begin{abstract}
Conjugated polymers are experiencing a surge of renewed interest due
to their promising applications in various organic electronic
devices. These include organic light-emitting diodes (OLEDs),
field-effect transistors (FETs), and organic photovoltaic (OPV)
devices, among many others. Their appeal stems from distinct
advantages they hold over traditional inorganic
semiconductors. Unlike inorganic semiconductors, where electrons are
often considered to be in delocalized, free, or quasi-free states
(as described by Bloch's theory), electrons in conjugated polymers
behave differently. They are strongly coupled within highly localized
$\sigma$ or $\pi$-orbitals and interact significantly with the ionic
cores. This means they are far from the idealized delocalized states
presumed by Bloch's theory approaches. Consequently, after nearly a
century of applying Bloch's theory to the electronic transport
properties of inorganic materials, there is a clear need for a new
theoretical framework to explain efficient charge transport in these
organic solids. Our presented model addresses this need by
incorporating crucial electron-electron interactions. Specifically,
it accounts for both intra-site interactions and interactions
between the $\pi$-states located at alternating sites along the polymer
chain. This framework provides a many-body charge conduction mechanism
and explains the semiconducting properties of the undoped material. A
significant outcome of our model is the prediction of two novel flat
bands of excited bonding states. Intriguingly, these states obey
Bose--Einstein statistics and facilitate charge transport. Furthermore,
our model accurately reproduces experimental data, providing an excellent
fit for measured UV-Vis absorption and electroluminescent spectra.
\end{abstract}


\maketitle

\section{Introduction}
\label{sec:intro}

Conjugated polymers (CPs) have attracted much attention for a long
time, both because of the interesting fundamental physics involved and
for their practical applications~\cite{LeKimYoon,Estarellas2017},
which even includes materials with promising use in neuromorphic
computation~\cite{Liu2023}. From the theoretical point of view the
work by Su, Schrieffer and Heeger~\cite{Su1,Su2,Heeger} did set an
important milestone. However, the interest on the subject matter was
revived recently because of their potential uses. The favorable
features of CPs are their easy synthesis and convenient processing,
diversity and flexibility. In the CPs the electrons are paired in
covalent highly localized $\sigma$-- or $\pi$--orbitals which interact
strongly between them and with the ionic cores. One of the aims of
this paper is to include, in the theoretical treatment of conjugated
polymers, the neighboring sites strong $\pi-\pi$ repulsion, which has
proven to be quite significant~\cite{Zhang}.

In contrast with the importance of their present and potential
applications~\cite{LeKimYoon}, the underlying physics of the
semiconducting and conducting properties of conjugated polymers is
still at an early stage of understanding. Compared to inorganic
semiconductors relatively little is known about the physical origin of
the electronic properties of the CPs, and even the precise nature of
the semiconductor excitations remains uncertain. Inorganic
semiconductors are characterized by the long range single crystal
spatial order, where delocalized non--interacting quasi--free
electrons in the periodic field of the crystalline lattice provide a
reasonable description. In contrast, in organic semiconductors the
strict spatial order is often reduced to one dimension, since the
different polymeric chains rarely form ordered arrays. More important,
the electrons are paired in covalent highly localized $\sigma$-- or
$\pi$--orbitals which interact strongly among themselves and with the
ionic cores, and quasi--free states do not provide a an adequate
treatment. Therefore, Bloch's theory for the electronic transport properties
of inorganic materials cannot be applied to the CPs, and one has to look
for a different theoretical framework for charge transport, since
quasi--free Bloch functions are not adequate to characterize the charge
carriers, since the strong repulsion between the $\pi$--electrons plays a
central role.

CPs are made up by identical molecular chain structures, held together
by a sequence of alternate single and double covalent bonds. The
double bonds combine a $\sigma$-- and a $\pi$--bond, whereas the
single ones are $\sigma-\sigma$--bonds. Hence the backbone is
essentially the more stable uniform sequence of $\sigma$--bonds. The
simplest CP is polyacetylene, for which the periodic molecular
structure is just a carbon atom with one of its valence electrons
bonded to a hydrogen atom. The metallic state is achieved by doping
through local oxidation or reduction. CP thin films attract most
of the attention, because of their many present and potential
applications in electronic and optoelectronic devices, such as
transistors, photodiodes, organic photovoltaic (OPV) devices, organic
light-emitting diodes (OLEDs) and many others, in addition to their
versatility, and both simple and low fabrication cost. Even the use of
paper as a substrate for organic transistors has been a subject of
investigation~\cite{Zschieschang}.

Some decades ago, devices that combine injected electrons from one
electrode and holes from the other one of a two--layer organic
emitting diode, were found to have good electroluminescent efficiency
under conveniently low operation voltages
\cite{Tang,Adachi,Burroughes,Braun}.  At present, the use of organic
materials as active semiconductors in electronic flat panel displays
are in large scale production and commercialization. CPs are
intrinsically stable under excitation by an applied voltage or photon
capture, both in light emission and harvesting devices. The
alternating $\pi$--bonds participate actively in the electronic
processes leaving intact the primary uniform backbone $\sigma$--bonded
structure. This backbone provides the necessary stability against
degradation by the energy transfer demanded in device operations.

In general, the carbon atoms of the backbone take electrons from
neighbouring atoms to form pairs with their four valence electrons in
order to create particularly stable octets. Hence the spatial
electronic pairing inherent to the $\pi$--bonds is taken here as a
primary property of conjugate polymers. Charge transport along the
polymer chain models must incorporate this fact, and presume that the
occurrence of any one--electron elementary process demands an
activation energy that is large. Bearing this in mind, we propose an
essentially many--body charge transport mechanism which meets the
conditions posed by the structure of the conjugated polymers, in a
more specific way than the conventional scheme, which was developed
for inorganic conductors and semiconductors. Our results are in good
agreement with UV--Vis photon absorption
experiments~\cite{Chen,Yao,Mangalore,Li,Lee,Park}.

\section{Model}
\label{sec:model}

The model put forward here constitutes a generalization of the ideas
pioneered by Su, Schrieffer and Heeger~\cite{Su1,Su2,Heeger}, by
adding explicitly the relevant electrostatic repulsion between
$\pi$--orbitals on neighbouring sites~\cite{Zhang}. Chain dimerization
is taken for granted and incorporated as an implicit
condition. Conduction state charge carriers have only subtle
differences with the solitons of Su, Schrieffer and Heeger. Our model
Hamiltonian for the polymer chain is

\begin{equation}
H=H_0+H_{\text{C0}}+H_{\text{C1}}+H_{\text{T}} ,
\label{E1}
\end{equation}
\noindent
where
\begin{equation}
H_0=\sum_l\epsilon\,\big( n_{l\uparrow}+n_{l\downarrow}\big) ,
\label{E2}
\end{equation}

\noindent
with $\epsilon$ denoting the energy per electron of the two
electronsof the $\pi$--orbital and $n_{l\uparrow}=c_{l\uparrow}^\dagger
c_{l\uparrow}$, $n_{l\downarrow}=c_{l\downarrow}^\dagger
c_{l\downarrow}$, are the two occupation operators of the one--electron
states at sites $l=-N/2,-N/2+1,\dots ,N/2$. The $H_{\text{C0}}$ term is

\begin{equation}
H_{\text{C0}}=\sum_j U_0 n_{j\uparrow}n_{j\downarrow},
\label{E3}
\end{equation}

\noindent
describes the Coulomb repulsion between electrons in $\pi$--orbitals
located on the same site of the polymer chain. Zhang et al.~\cite{Zhang}
proved that the $\pi$--$\pi$ electrostatic repulsion in conjugated
systems has also significant effects in neighbouring sites, while the
$\pi$--$\pi$ Pauli repulsion plays a secondary but non--negligible role.
The term

\begin{equation}
H_{\text{C1}}=
\sum_l U_1 \big(n_{l+1\uparrow}+n_{l+1\downarrow}\big)
\big(n_{l\uparrow}+n_{l\downarrow}\big).
\label{E4}
\end{equation}

\noindent
accounts for this effect. The two--electron tunneling term

\begin{equation}
H_{\text{T}}=
\sum_l V\big(c_{l+1\uparrow}^\dagger c_{l+1\downarrow}^\dagger
c_{l\downarrow}c_{l\uparrow}
+c_{l\uparrow}^\dagger c_{l\downarrow}^\dagger
c_{l+1\downarrow}c_{l+1\uparrow}\big)
\label{E5}
\end{equation}

\noindent
accounts for quantum fluctuations of the $\pi$--bonds between
neighboring sites. The underlying principle is that the low energy
$N$--electron states of the chain can be expressed as combinations of
two--electron $\pi$--states localized on every other site along the
chain. Hence, any term of the Hamiltonian not having this general
structure yields a vanishing contribution when operating on the paired
states.

\subsection{Transformation to spin-1/2 operators}

In the following sections we will use the just introduced Fermi--Dirac
electron representation. However, a transformation to spin 1/2
operators may be also be an advantage. Introducing  the new 
dynamical variables as

\begin{equation}
\begin{aligned}
& s_1(l)=\frac12\big( c_{l\uparrow}^\dagger c_{l\downarrow}^\dagger
+c_{l\downarrow}c_{l\uparrow}\big)\\
& s_2(l)=\frac1{2i}\big( c_{l\uparrow}^\dagger c_{l\downarrow}^\dagger
-c_{l\downarrow}c_{l\uparrow}\big)\\
& s_3(l)=\frac12\big( n_{l\uparrow}+n_{l\downarrow}-1\big) \ ,
\end{aligned}
\label{E6}
\end{equation}
which satisfy the usual angular momentum commutation relations
\begin{equation}
[s_1,s_2]=is_3 \quad [s_2,s_3]=is_1 \quad [s_3,s_1]=is_2\ ,
\label{E7}
\end{equation}
and the Hamiltonian $H$ takes the general form of the Hamiltonian of an
anisotropic spin 1/2 antiferromagnetic Heisenberg model

\begin{equation}
\begin{aligned}
H=\,&\sum_l\big[\epsilon+2U_0 +2U_1 +U_0 s_3(l)\big]\\
&+4U_1\sum_l\bigg( s_3(l+1)s_3(l)\\
&+\frac{V}{2U_1}\big[ s_1(l+1)s_1(l)+s_2(l+1)s_2(l)\big]\bigg)\ .
\label{E8}
\end{aligned}
\end{equation}

\noindent
As the number operators $c_l^\dagger c_l$ have eigenvalues 0 and 1,
the eigenvalues of $s_3$ are $-1/2$, $0$ and $1/2$. The eigenvalue $0$
describes the breaking of a covalent bond, which has a large energy
cost. In the limit it becomes infinite the operators $s_1$,
$s_2$, and $s_3$ behave as the components of a spin 1/2.

In terms of the ladder operators $s_+=s_1+is_2$ and $s_-=s_1-is_2$ $H$
can be rewritten as

\begin{equation}
\begin{aligned}
H=\,&\sum_l\big[\epsilon+2U_0 +2U_1 +U_0 s_3(l)\big]\\
&+4U_1\sum_l\bigg( s_3(l+1)s_3(l)\\
&+\frac{V}{4U_1}\big[ s_{+}(l+1)s_{+}(l)+s_{-}(l+1)s_{-}(l)\big]\bigg)\ .
\label{E9}
\end{aligned}
\end{equation}

\noindent
Since only full occupation is considered, the first sum on the right
hand side of Eq.~(\ref{E9}) is a constant, and can be ignored.
The use of fermion or angular momentum operators are two formally
equivalent alternatives to deal with the model put forward here.

\section{Ground state and ground state energy}
\label{GroundState}
\subsection{The ladder operators}
\label{ladder}

Techniques established for handling the anisotropic Heisenberg
antiferromagnetic chain can be effectively applied by translating
these systems into the fermion scheme. This approach
\cite{LagosCabrera,LagosKiwiGaglianoCabrera,CabreraLagosKiwi,
GottliebLagosMontenegro}, starts defining {\it even} and {\it odd}
operators within the fermion framework.

\begin{equation}
\begin{aligned}
\phi_{\text{e}}^\dagger &=&\sqrt{\frac{2}{N}}
\sum_{\text{even }l} c_{l+1\uparrow}^\dagger c_{l+1\downarrow}^\dagger
c_{l\downarrow}c_{l\uparrow}+
\frac{\alpha}{2}\sqrt{\frac{N}{2}},
\label{E10}\\
\phi_{\text{o}}^\dagger &=&\sqrt{\frac{2}{N}}
\sum_{\text{odd }l} c_{l\uparrow}^\dagger c_{l\downarrow}^\dagger
c_{l+1\downarrow}c_{l+1\uparrow}+
\frac{\alpha}{2}\sqrt{\frac{N}{2}},
\end{aligned}
\end{equation}

\noindent
where $\alpha =V/2U_1$. This sub--section is devoted to derive the main
properties of these operators, both the exact ones and their asymptotic
limit for strong repulsion between the $\pi$--states. Their relevance to
the present problem will become apparent in the next subsection.

Recalling the elementary identities between commutators and
anticommutators $[A,BC]_-= [A,B]_-C+B[A,C]_-$,
$[AB,C]_-= A[B,C]_-+[A,C]_-B$, $[A,BC]_-=\{A,B\}_+C-B\{A,C\}_+$ and
$[AB,C]_-=A\{B,C\}_+-\{A,C\}_+B$, where $\{ \}_+$ and $[ ]_-$ correspond
to anti-commutators and commutators, respectively, and otherwise the
usual notation is employed. One can now show the following properties of the
$\phi$--operators

\begin{equation}
\begin{aligned}
&[\phi_{\text{e}},\,\phi_{\text{e}}^\dagger ]
=\frac{2}{N}\sum_{\text{even }l}
\big[ n_{l\uparrow} n_{l\downarrow}
\big( 1-n_{l+1\uparrow}-n_{l+1\downarrow} \big)\\
&\phantom{abababababab}-n_{l+1\uparrow} n_{l+1\downarrow}
\big( 1-n_{l\uparrow}-n_{l\downarrow} \big)\big],
\label{E11}
\end{aligned}
\end{equation}

\begin{equation}
\begin{aligned}
&[\phi_{\text{o}},\,\phi_{\text{o}}^\dagger ]
=-\frac{2}{N}\sum_{\text{odd }l}
\big[ n_{l\uparrow} n_{l\downarrow}
\big( 1-n_{l+1\uparrow}-n_{l+1\downarrow} \big)\\
&\phantom{ababababababa}-n_{l+1\uparrow} n_{l+1\downarrow}
\big( 1-n_{l\uparrow}-n_{l\downarrow} \big)\big],
\label{E12}
\end{aligned}
\end{equation}

\begin{equation}
[\phi_{\text{e}},\,\phi_{\text{o}}]\equiv 0,
\label{E13}
\end{equation}

\begin{equation}
\begin{aligned}
&\bigg[\sum_l n_{l\uparrow}n_{l\downarrow}, \phi_e^\dagger \bigg]
=\sqrt{\frac{2}{N}}\sum_{\text{even }l}
c_{l+1\uparrow}^\dagger c_{l+1\downarrow}^\dagger
c_{l\downarrow}c_{l\uparrow}(1-n_{l\uparrow}-n_{l\downarrow})\\
&\phantom{ababababab}+\sqrt{\frac{2}{N}}
\sum_{\text{odd }l}c_{l\uparrow}^\dagger c_{l\downarrow}^\dagger
c_{l-1\downarrow}c_{l-1\uparrow}(1+n_{l\uparrow}+n_{l\downarrow}),
\label{E14}
\end{aligned}
\end{equation}

\begin{equation}
\begin{aligned}
&\bigg[\sum_l n_{l\uparrow}n_{l\downarrow}, \phi_o^\dagger \bigg]
=\sqrt{\frac{2}{N}}\sum_{\text{even }l}
c_{l-1\uparrow}^\dagger c_{l-1\downarrow}^\dagger
c_{l\downarrow}c_{l\uparrow}(1-n_{l\uparrow}-n_{l\downarrow})\\
&\phantom{ababababba}+\sqrt{\frac{2}{N}}\sum_{\text{odd }l}
c_{l\uparrow}^\dagger c_{l\downarrow}^\dagger
c_{l+1\downarrow}c_{l+1\uparrow}(1+n_{l\uparrow}+n_{l\downarrow}),
\label{E15}
\end{aligned}
\end{equation}

\begin{equation}
\begin{aligned}
&[H_{C1},\,\phi_{\text{e}}^\dagger]=
2U_1\sqrt{\frac{2}{N}}\sum_{l\text{ even}}
c_{l+1\uparrow}^\dagger c_{l+1\downarrow}^\dagger
c_{l\downarrow}c_{l\uparrow}\\
&\phantom{abababab}\times (n_{l+2\uparrow}+n_{l+2\downarrow}
-n_{l+1\uparrow}-n_{l+1\downarrow}\\
&\phantom{abababab}+n_{l\uparrow}+n_{l\downarrow}-n_{l-1\uparrow}
-n_{l-1\downarrow}-2),
\label{E16}
\end{aligned}
\end{equation}

\begin{equation}
\begin{aligned}
&[H_{C1},\,\phi_{\text{o}}^\dagger]=
2U_1\sqrt{\frac{2}{N}}\sum_{l\text{ odd}}
c_{l\uparrow}^\dagger c_{l\downarrow}^\dagger
c_{l+1\downarrow}c_{l+1\uparrow}\\
&\phantom{abababab}\times (-n_{l+2\uparrow}-n_{l+2\downarrow}
+n_{l+1\uparrow}+n_{l+1\downarrow}\\
&\phantom{abababab}-n_{l\uparrow}-n_{l\downarrow}+n_{l-1\uparrow}
+n_{l-1\downarrow}-2).
\label{E17}
\end{aligned}
\end{equation}
and

\begin{equation}
H_{\text{T}}=
\sqrt{\frac{N}{2}}V(\phi_e^\dagger +\phi_o^\dagger +\phi_e +\phi_o)
-N\alpha V.
\label{E18}
\end{equation}

In the asymptotic strong conjugation limit

\begin{equation}
n_{l\uparrow}=n_{l\downarrow}\rightarrow
\begin{cases}
1,\text{ if }l\text{ even}\\
0,\text{ if }l\text{ odd}
\end{cases}
\label{E19}
\end{equation}
and the commutators in Eqs.~(\ref{E11}) and (\ref{E12}) yield the Bose
commutation relations

\begin{equation}
[\phi_{\text{e}},\,\phi_{\text{e}}^\dagger ]
=[\phi_{\text{o}},\,\phi_{\text{o}}^\dagger ]=1 .
\label{E20}
\end{equation}
Also, in the same strong conjugation limit, for long enough polymeric
chains one can write 

\begin{equation}
[H_{C0},\phi_{\text{e,o}}^\dagger ]=0\ ,
\label{E21}
\end{equation}
and all terms in the two summations on the right-hand sides of
Eqs.~(\ref{E14}) and (\ref{E15}) cancel in the limit of
Eq.~(\ref{E19}), except for their first and last terms. A similar
argument applies to $H_0$, and in the limit of Eq.~(\ref{E19}) it also holds,
and yields
\begin{equation}
[H_{C1},\phi_{\text{e,o}}^\dagger ]=
4U_1\bigg(\phi_{\text{e,o}}^\dagger
-\frac{\alpha}{2}\sqrt{\frac{N}{2}}\bigg)\ .
\label{E22}
\end{equation}
Therefore, in the strong conjugation limit of Eq.~(\ref{E19}) one
obtains from Eqs.~(\ref{E1}), (\ref{E18}), (\ref{E20}), (\ref{E21})
and Eq.~(\ref{E22}) that

\begin{equation}
[H,\phi_{\text{e}}^\dagger ]=4U_1\phi_{\text{e}}^\dagger ,\qquad
[H,\phi_{\text{o}}^\dagger ]=4U_1\phi_{\text{o}}^\dagger .
\label{E23}
\end{equation}
Hence, in the limit of strong conjugation, the operators
$\phi_{\text{e}}$ and $\phi_{\text{o}}$ are ladder operators, which
transform any stationary state of $H$ into a state of lower energy.
Operators $\phi_{\text{e}}^\dagger$ and $\phi_{\text{o}}^\dagger$ have a
similar effect, but change the states to higher energy ones. Therefore the
ground state $|g\rangle$ of $H$ must satisfy

\begin{equation}
\phi_{\text{e}}|g\rangle =\phi_{\text{o}}|g\rangle =0 \ .
\label{E24}
\end{equation}
because no stationary state of lower energy can exist.

\subsection{The ground state $|g\rangle$}

Defining $|\mathcal{N}\rangle$ as the chain bare of $\pi$--electronic
states (in the spin representation $|\mathcal{N}\rangle$ is the N{\'e}el
state)

\begin{equation}
|\mathcal{N}\rangle =\prod_{\text{even }l}
c_{l\uparrow}^\dagger c_{l\downarrow}^\dagger |0\rangle ,
\label{E25}
\end{equation}

\noindent
where $|0\rangle$ is the vacuum, that is the naked backbone of
$\sigma$--bonds, and introducing the operator $\Lambda$ as 

\begin{equation}
\begin{aligned}
\Lambda &=\sqrt{\frac{N}{2}}\big(\phi_{\text{e}}^\dagger
+\phi_{\text{o}}^\dagger -\phi_{\text{e}}-\phi_{\text{o}}\big)\\
&=\sum_l (-1)^l \big(
c_{l+1\uparrow}^\dagger c_{l+1\downarrow}^\dagger
c_{l\downarrow}c_{l\uparrow}
-c_{l\uparrow}^\dagger c_{l\downarrow}^\dagger
c_{l+1\downarrow}c_{l+1\uparrow} \big)\ ,
\label{E26}
\end{aligned}
\end{equation}

\noindent
it is shown next that in the asymptotic limit of Eq.~(\ref{E19}) the ground
state of $H$ is given by

\begin{equation}
|g\rangle
=\exp\bigg(-\frac{\alpha}{2}\Lambda\bigg)\ |\mathcal{N}\rangle .
\label{E27}
\end{equation}

To demonstrate the validity of Eq.~(\ref{E27}) notice that it can be
proven by complete induction that, in the strong conjugation limit of
Eq.~(\ref{E19}), it yields

\begin{equation}
[\phi_{\text{e,o}},
(\phi_{\text{e}}^\dagger +\phi_{\text{o}}^\dagger -\phi_{\text{e}}
-\phi_{\text{o}})^n]
=n(\phi_{\text{e}}^\dagger +\phi_{\text{o}}^\dagger -\phi_{\text{e}}
-\phi_{\text{o}})^{n-1}
\label{E28}
\end{equation}
for $n=0,1,2,\dots$, hence for any analytic function $F$ with derivative
$F^\prime$ 

\begin{equation}
[\phi_{\text{e,o}},F(\phi_{\text{e}}^\dagger
+\phi_{\text{o}}^\dagger -\phi_{\text{e}}
-\phi_{\text{o}})]=F^\prime (\phi_{\text{e}}^\dagger
+\phi_{\text{o}}^\dagger -\phi_{\text{e}}
-\phi_{\text{o}}).
\label{E29}
\end{equation}
Applying this property with $F$ substituted by the exponential
function appearing in Eq.~(\ref{E27}) and the definitions of
Eq.~(\ref{E10}), it can readily be shown that $|g\rangle$ satisfies
Eqs.~(\ref{E24}) and is therefore the ground state in the strong
conjugation limit Eq.~(\ref{E19}). Notice that the ground state
Eq.~(\ref{E27}) is not perturbative because of the large factor
$\sqrt{N/2}$ multiplying the sum Eq.~(\ref{E26}) that defines
$\Lambda$.

\subsection{The ground state energy $E_g$}
\label{gs_energy}

To determine the ground state energy the commutation property
of $\Lambda$
\begin{equation}
\begin{aligned}
&[c_{l\uparrow}^\dagger c_{l\downarrow}^\dagger ,\Lambda]\\
&=(-1)^l\big( c_{l+1\uparrow}^\dagger c_{l+1\downarrow}^\dagger
+c_{l-1\uparrow}^\dagger c_{l-1\downarrow}^\dagger\big)
\big( n_{l\uparrow}+n_{l\downarrow}-1\big),
\end{aligned}
\label{E30}
\end{equation}
is required, which  in the asymptotic limit Eq.~(\ref{E19}), after
iterating $\nu$ times, it reads 

\begin{equation}
\begin{aligned}
&[[\dots [c_{l\uparrow}^\dagger c_{l\downarrow}^\dagger ,\Lambda],
\Lambda ],\dots ,\Lambda]_{\nu\text{ times}}\\
&\phantom{abab}
=(-1)^{l\nu}(-1)^{\nu (\nu +1)/2}\big(\tau +\tau^{-1}\big)^{\nu}
c_{l\uparrow}^\dagger c_{l\downarrow}^\dagger ,
\end{aligned}
\label{E31}
\end{equation}
where $\tau$ is the nearest neighbor translation operator

\begin{equation}
\tau c_{ls}=c_{l+1s}\quad\tau^{-1} c_{ls}=c_{l-1s}\quad
s=\uparrow ,\downarrow .
\label{E32}
\end{equation}
Combining this with the identity

\begin{equation}
\begin{aligned}
e^{-B}A\, e^B \equiv &A+\frac{1}{1!}[A,B]+\frac{1}{2!}[[A,B],B]\\
&+\frac{1}{3!}[[[A,B],B],B]+\cdots ,
\label{E33}
\end{aligned}
\end{equation}
the generating function of the modified Bessel functions $I_\nu (z)$

\begin{equation}
\begin{aligned}
\exp\bigg[\frac{z}{2}\big(\tau +\tau^{-1}\big)\bigg]
=\sum_{\nu =-\infty}^\infty I_\nu (z)\tau^\nu ,
\label{E34}
\end{aligned}
\end{equation}
and the property $J_\nu (z)=i^{-\nu}I_\nu(iz)$, where $J_\nu$ is the
unmodified Bessel function, it can be shown that

\begin{equation}
\begin{aligned}
&\exp\bigg(\frac{\alpha}{2}\Lambda\bigg)
c_{l\uparrow}^\dagger c_{l\downarrow}^\dagger
\exp\bigg(-\frac{\alpha}{2}\Lambda\bigg)\\
&=\sum_{\nu =-\infty}^\infty (-1)^{l\nu} (-1)^{\nu (\nu +1)/2}
J_\nu (\alpha )\, c_{l+\nu\uparrow}^\dagger
c_{l+\nu\downarrow}^\dagger .
\label{E35}
\end{aligned}
\end{equation}

Eq.~(\ref{E35}) in combination with Eq.~(\ref{E27}) are now used to
calculate expectation values. In particular, the average occupation of
a site $l$ is

\begin{equation}
\langle g|(n_{l\uparrow}+n_{l\downarrow})|g\rangle
= 1+(-1)^l J_0(2\alpha ) \ ,
\label{E36}
\end{equation}
where $J_\nu$ is the usual Bessel function of order $\nu $. The short
range correlation coefficient is 

\begin{equation}
\frac{1}{N}\langle g|
\sum_l\big( n_{l+1\uparrow}+n_{l+1\downarrow}\big)
\big( n_{l\uparrow}+n_{l\downarrow}\big) |g\rangle
= 1-[J_0(2\alpha )]^2 .
\label{E37}
\end{equation}
The energy of the ground state $E_g = \langle g|\big( H_0+H_{C1}+H_T\big)
|g\rangle $ is given by

\begin{equation}
E_g =N\epsilon +NU_1\big[1-\big( J_0(2\alpha )\big)^2 +2\alpha J_1(2\alpha )\big] \ .
\label{E38}
\end{equation}
In obtaining Eqs.~(\ref{E36}), (\ref{E37}) and Eq.~(\ref{E38}) use was
made of Neumann's addition formulas of the Bessel functions and Graf's
generalization of them~\cite{Watson}. The previous results expressed
in terms of the Bessel functions look elegant, but it must be kept in
mind that they hold only in the asymptotic limit Eq.~(\ref{E19}), that is when
$\alpha$ is sufficiently small. Up to second order in $\alpha$ one has that

\begin{equation}
E_g =4NU_1(\alpha^2 +0(\alpha^4))\ .
\label{E39}
\end{equation}

Using the properties of the ladder operators $\phi_{\text{e}}^\dagger$
and $\phi_{\text{o}}^\dagger$, given by Eq.~(\ref{E23}), yields the set of
eigenvectors 

\begin{equation}
|n_{\text{e}}\, n_{\text{o}}\rangle
=\dfrac{(\phi_{\text{e}}^\dagger)^{n_{\text{e}}}}{\sqrt{n_{\text{e}}!}}
\dfrac{(\phi_{\text{o}}^\dagger)^{n_{\text{o}}}}{\sqrt{n_{\text{o}}!}}
\,|g\rangle , \quad
n_{\text{e}},\, n_{\text{o}}=0,1,2,3,\dots
\label{E40}
\end{equation}
which are eigenvectors of $H$, with eigenenergies

\begin{equation}
E_{n_{\text{e}}n_{\text{o}}}=4(n_{\text{e}}+n_{\text{o}})U_1+E_g.
\label{E41}
\end{equation}

The theoretical framework described up to this point would not be
complete without including the fact that the ground state is twofold
degenerate. In effect,
\begin{equation}
|\bar{g}\rangle
=\exp\bigg(\frac{\alpha}{2}\Lambda\bigg)\ | \bar{\mathcal{N}}\rangle,
\quad | \bar{\mathcal{N}}\rangle =\prod_{\text{odd }l}
c_{l\uparrow}^\dagger c_{l\downarrow}^\dagger |0\rangle ,
\label{E42}
\end{equation}
is also an eigenvector of $H$ with the same energy eigenvalue
Eq.~(\ref{E38}). The bosonic operators $\phi_{\text{e}}$ and
$\phi_{\text{o}}$ become creation operators when operating on
$|\bar{g}\rangle$.

\begin{figure}[h!]
\begin{center}
\includegraphics[width=75mm]{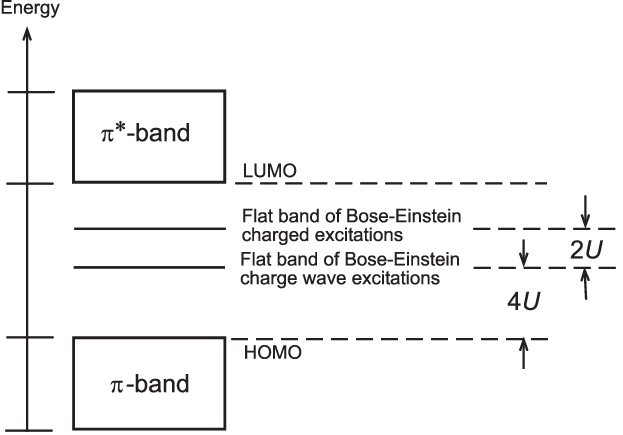}
\caption{\label{Fig1} The $\pi$--$\pi$ repulsion gives rise to two
  Bose--Einstein excitations energy levels, whose energies lie inside
  the gap $4U$ and $6U$ above the energy of the highest occupied
  molecular orbital (HOMO).}
\end{center}
\end{figure}

Fig.~\ref{Fig1} shows a schematic diagram of the energy spectrum of
the molecular chain. The $\pi$--$\pi$ repulsion yields a novel energy
level of Bose--Einstein excitations located between the HOMO and
LUMO energies. Hence, the conduction band in this framework is not
made up by electrons in extended anti--bonding states, but by
electronic pairs excited to the novel flat band of energy $4U$.
Certainly the system has other degrees of freedom, additional to the
ones described by the operators $\phi_{\text{e}}$ and
$\phi_{\text{o}}$, e.g.~the solitons first invoked by Su, Schrieffer and
Heeger~\cite{Su1,Su2} for polyacetilene, and later generalized to other
conjugate polymers~\cite{Heeger}. They give rise to the flat band of
energy $6U_1$ in Fig.~\ref{Fig1}, which we will address in detail, after
we show how the excitations associated with the operators
$\phi_{\text{e}}$ and $\phi_{\text{o}}$ transport momentum and energy. As
it is assumed that one--electron processes involve large activation
energies, the momentum operator $\vec{P}$ must be written as the
two--particle operator

\begin{equation}
\begin{aligned}
\vec{P}=
& -\frac{i\hbar}{2}\int d^3{r'}\ d^3 {r}\, \Psi^\dagger(\vec{r'},t)
\Psi^\dagger(\vec{r},t) \\
&\times (\nabla' +\nabla)\Psi(\vec{r},t)\Psi(\vec{r'},t)
+\,\text{adjoint operator},
\label{E43}
\end{aligned}
\end{equation}

and in terms of the electron field operator it reads

\begin{equation}
\begin{aligned}
\Psi(\vec{r},t)=\sum_{ls}c_{ls} w_s(\vec{r}-la\hat{\imath}).
\label{E44}
\end{aligned}
\end{equation}

\noindent
Here $\nabla'$ and $\nabla$ are the gradient operators with respect to
$\vec{r'}$ and $\vec{r}$, and $w_s(\vec{r}-la\hat{\imath})$ represents
the one--particle wave function of an electron in a covalent $\pi$ state
at site $l$, $a$ is the distance between nearest neighbor sites of the
polymer chain, and $\hat{\imath}$ is the unitary vector along the chain
direction. Because of analytical reasons, and the small overlap of
functions $w_s$ centered on neighboring sites 

\begin{equation}
\int d^3 {r}\, w_s^*(\vec{r}-l'a\hat{\imath})
\nabla w_s(\vec{r}-la\hat{\imath}) =
\begin{cases}
0,\text{ if } l=l'\\
q\,\hat{\imath}\text{ if } l'=l+1\\
-q^*\,\hat{\imath}\text{ if } l'=l-1\\
0, \text{ otherwise} \ .
\end{cases}
\label{E45}
\end{equation}
Inserting Eqs.~(\ref{E44}) and (\ref{E45}) into Eq.~(\ref{E43}), the
resulting expression for $\vec{P}$ finally reduces to the same
standard equation for the one--particle momentum operator.  However,
one must recall that the polymer backbone is not rigid, and the
alternating occupied and unoccupied $\pi$--orbitals cause a
dimerization of the polymer chain. Hence, the occupied and virtual
$\pi$--orbitals are expected to have finite energy differences, simply
because of the broken periodicity of the distance between the positive
charges involved in the chemical bonds. In Eq.~(\ref{E45}) the
parameter $a$ takes slightly different values if the accompanying
index $l$ is even or odd. When taking this into consideration the
momentum operator splits into a one--particle and a two--particle
term, and takes the general form
\begin{equation}
\begin{aligned}
& \vec{P}=-i\hbar\,q\,\hat{\imath}
\sum_{l,s}\big( c_{l+1,s}^\dagger c_{ls}
-c_{l,s}^\dagger c_{l+1,s}\big)\\
& -i\hbar\,\gamma q\,\hat{\imath}
\sum_l\big( c_{l+1\uparrow}^\dagger c_{l+1\downarrow}^\dagger
c_{l\downarrow}c_{l\uparrow}
-c_{l\uparrow}^\dagger c_{l\downarrow}^\dagger
c_{l+1\downarrow}c_{l+1\uparrow}\big),
\label{E46}
\end{aligned}
\end{equation}

\noindent
where $\gamma$ is a coefficient proportional to the shift $\delta a$ of
the bond lengths of the dimerized chain. In general, the first term of
$\vec{P}$ in Eq.~(\ref{E46}) destroys pairs and the second one always
conserves them. As $H$ and the eigenstates Eq.~(\ref{E27}), Eq.~(\ref{E40}) and
Eq.~(\ref{E42}) involve just paired electrons, consistent with the principle
that one--electron processes involve too large activation energies, it is
sufficient to retain just the second term and write $\vec{P}$ as

\begin{equation}
\begin{aligned}
\vec{P} & =-i\hbar\,\gamma q\,\hat{\imath}
\sqrt{\frac{N}{2}}\big(\phi_{\text{e}}^\dagger
-\phi_{\text{o}}^\dagger -\phi_{\text{e}}+\phi_{\text{o}}\big)\ .
\label{E47}
\end{aligned}
\end{equation}

In the Heisenberg picture
\begin{equation}
\frac{d^2P}{dt^2}=-\frac{1}{\hbar^2}[H,[H,P]] 
\label{E48}
\end{equation}
and using Eqs.~(\ref{E47}) and (\ref{E23}) one has that

\begin{equation}
\frac{d^2P}{dt^2} = -\frac{4U_1^2}{\hbar^2}P \ .
\label{E49}
\end{equation}

\noindent
Eq.~(\ref{E49}) shows that the operators $\phi_{\text{e}}^\dagger$ and
$\phi_{\text{o}}^\dagger$ excite collective charge motion modes of the
chain $\pi$--bonds. The collective oscillation involves charge
displacements of the angular frequency $\omega =2U_1/\hbar$, independent
of the polymer chain length. Hence, resonances favouring charge
transfer with neighbouring chains are expected to occur.

The model worked out above shows that the dynamical effect of the
$\pi$--$\pi$ repulsion in conjugate polymers is a set of degenerate
excited states, which can transport charge along the polymer
chain. These novel states obey Bose--Einstein statistics and have energy
eigenvalues that form a flat band located $4U_1$ above the ground state,
in the HOMO-LUMO gap. The model assumed in this step provides a natural
answer to the question of how the repulsion between the $\pi$--orbitals
affects the dynamics of the system, however it still is a bit too
simple. An important conclusion is that the single frequency
$\omega =2U_1/\hbar$, common to all the polymeric chains in the sample,
no matter their length, are expected to produce resonances that
enhance the probability of charge transfer.

\begin{figure}[h!]
\begin{center}
\includegraphics[width=80mm]{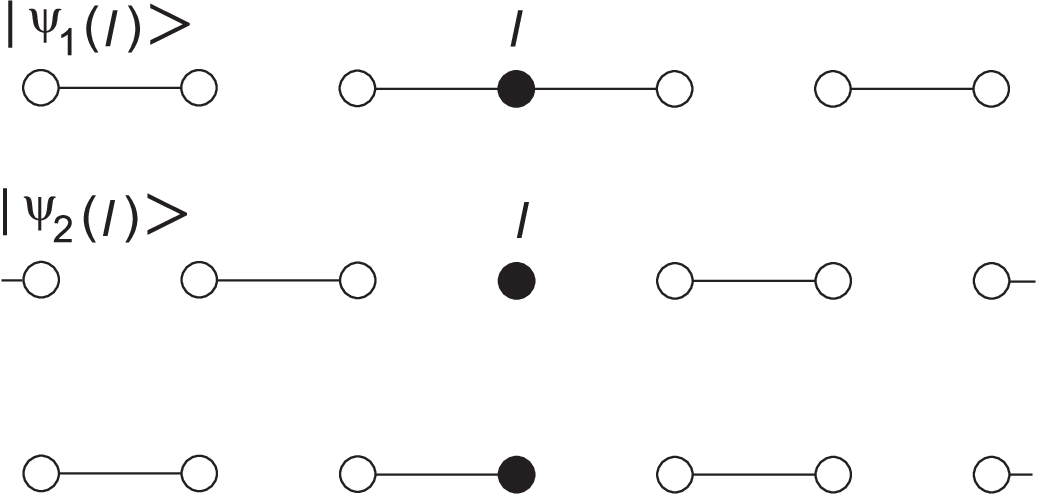}
\caption{\label{Fig2} Circles represent monomers and lines the
$\pi$--bonds. The filled black circles are for reference only. The upper
and middle illustrations represent the hybrid states
$|\psi_1\rangle$ and $|\psi_2\rangle$, respectively. The bottom
configuration represents one of the two degenerate ground states.}
\end{center}
\end{figure}

\subsection{Metallic phase}

The metallic phase is made up by a superposition of excited states,
quite similar to those introduced by Su et
al.~\cite{Su1,Su2,Heeger}. At any site $l$ the $\pi$--orbital
occupation changes from even to odd, or vice--versa. The monomer in
between marks a boundary between the two degenerate ground states,
$|g\rangle$ and $|\bar{g}\rangle$, that prevail at either side of the
monomer. The energy of this state can be inferred from a heuristic
argument. First consider a class of excited states of the form
$(|\psi_1(l) \rangle\pm |\psi_2(l)\rangle)/\sqrt{2}$, where
$|\psi_1(l)\rangle$ and $|\psi_2(l)\rangle$ are the states represented
schematically in Fig.~\ref{Fig2}. As the interaction between the
orbitals is limited to nearest neighbors, for $\alpha$ small enough
the mutual perturbation of the states at the two sides of $l$ is
small. 

Therefore, the combined energy contribution from both sides totals
approximately $4U_1$.  Additionally, the energy contribution from the
central parts at site $l$ of $|\psi_1(l)\rangle$ and
$|\psi_2(l) \rangle$ can be estimated in zeroth order in $\alpha$ as
$4U$ and 0, respectively.  Hence, the contribution to the energy of
the central part is the average value $2U$ between these two figures,
which added to the energy $4U$ of the two sides gives an estimate of
$6U_1$ for the total energy of the combined states
$(|\psi_1(l)\rangle\pm |\psi_2(l)\rangle)/ \sqrt{2}$. This class of
excited states configures the upper flat band of charged excitations
illustrated in Fig.~\ref{Fig1}. The excitations have a charge of $\pm 2e$
and an energy $6U_1$ above the energy of the highest occupied
molecular orbital (HOMO). We understand that the energy of the HOMO
incorporates the energy Eq.~(\ref{E38}) for the vacuum of the new
excitations. The metallic phase is given by the linear combination

\begin{equation}
\begin{aligned}
&|k\rangle =\sqrt{\frac{1}{N}}\sum_{l=-N/2}^{N/2}\exp(ikl)
|\psi_p(l)\rangle \ ,
\quad k=\frac{2\pi}{N}n,\\
&\quad p=1,2 \qquad n=-\frac{N}{2},\dots,-1,0,1,2,\dots,\frac{N}{2} . 
\end{aligned}
\label{E50}
\end{equation}

We now compute the experimentally measurable results that the above
model yields.

\section{Comparison with experiment}
\label{sec:experiment}

The predictions obtained with the above model for the energy bands can
readily be tested by light absorption spectra. The two prominent flat
bands observed in Fig.~\ref{Fig1} are particularly characteristic,
notably exhibiting an approximate energy ratio of 3/2. This
distinctive feature was examined by Chen et al.~\cite{Chen} through
their measurements of the electromagnetic radiation absorption spectra
of a 60-nm thick film of the conjugated polymer FBT-Th$_4$(1,4) in the
300--800 nm wavelength interval. The target polymer was a solid film;
however, solutions in chlorobenzene and dichlorobenzene gave very
similar results. Two main maxima, at $\lambda_1=692\text{ nm}$ and
$\lambda_2=453\text{ nm}$ were observed; their energies are
$\epsilon_1=1.792\text{eV}$ and $\epsilon_2=2.737\text{eV}$, and their
ratio is
\begin{equation}
\frac{\epsilon_2}{\epsilon_1}=1.527 ,
\label{E51}
\end{equation}
which is very close to the estimated $3/2$ value. The accuracy of the
agreement is quite unexpected, because the spectral maxima are
displaced from the actual transition energy by Stokes shifts; however,
the two maxima displacements are expected to be quite similar, which
may explain such a good agreement. The rest of the experimental
spectrum also is in good agreement with the energy level scheme of
Fig.~\ref{Fig1}; in fact, the maximum between a plateau and a zero
absorption region are evidence of an energy gap.

\begin{figure}[h!]
\begin{center}
\includegraphics[width=85mm]{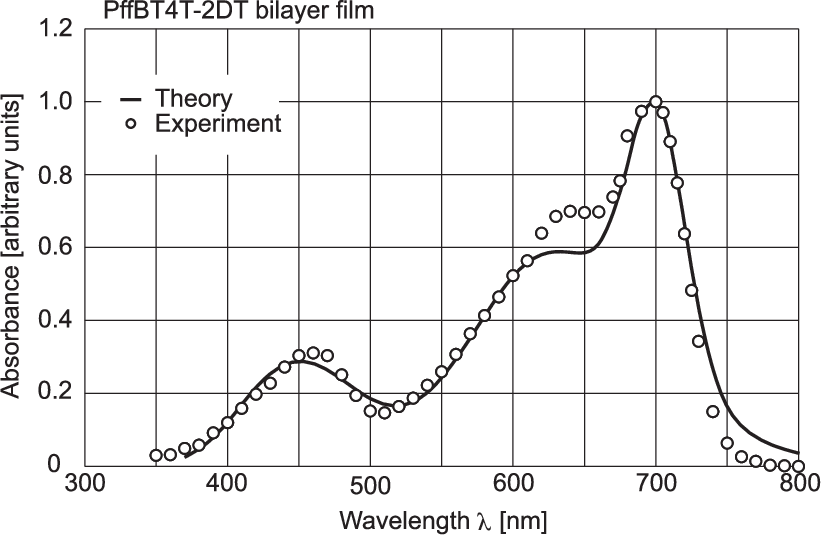}
\caption{\label{Fig3} Open circles illustrate experimental results for
  the UV--Vis bsorption spectra of a thin film of the organic
  semiconductor PffBT4T-2DT, a conjugate polymer~\cite{Mangalore}. The
  full line is a rough theoretical estimate of the processes, assuming
  the excitation of three sharp energy levels, that correspond (from
  left to right) to wavelengths 620, 830 and 752~nm for the three wide
  peaks. The large widths and shifts of the maxima are due to the
  multiphonon character of the processes. Just bulk substrate modes
  were considered in the estimate, which shows that the observed
  absorption bands can be explained on the basis of a few sharp energy
  levels.}
\end{center}
\end{figure}

More recent observations of the UV--Vis excitation spectra, of many
conjugated polymers of interest, show results similar to those of Chen
et al.~\cite{Chen}, with the ratio of Eq.~(\ref{E51}) ranging between
1.4 and 1.7~\cite{Yao,Mangalore,Li,Lee,Park}. The spectra shows two
main peaks, centered typically at $\lambda_1\approx 700\, \text{nm}$
and $\lambda_2\approx 450\,\text{nm}$, with a full width at half
maximum of about 100~nm, as illustrated in Fig.~\ref{Fig3}.

However, the most remarkable feature of the measured spectra is that
the absorption and emission bands can be explained by the excitation
and de--excitation of a few states of sharply defined energies, as
those predicted in the previous sections. A typical example is given
in Fig.~\ref{Fig3}, where the open circles represent the photon
absorbance of a thin layer of the semiconductor CP PffBT4T-2DT,
measured by Li et al.~\cite{Mangalore}. The solid line is a
theoretical estimate, which assumes that photon absorption takes place
by the excitation of three sharp levels of energies 2.00, 1.49 and
1.65~eV (wavelengths 620, 830 and 752 nm, respectively). The three
absorption lines are significantly widened and shifted by multiphonon
events that accompany the excitation process. The excitation of a
bonding orbital necessarily involves a sudden distortion of the
molecular environment, which excites the vibration modes, particularly
the acoustic ones. Since the entropy increases the vibration modes
exchange energy at random, but with an average energy gain. This
widens the spectral peaks and shifts them to shorter wavelengths
(Stokes shift) as they absorb photons. The 620~nm level becomes a wide
line that peaks around 450~nm, while the 830~nm level contributes the
largest intensity, and is significantly broadened giving rise to the
shoulder at $\sim$~650~nm. Finally, the 752~nm state is only slightly
broadened and manifests itself as the narrow peak at 700~nm.

The fit of the theoretical curve of Fig.~\ref{Fig3}, was obtained
following the procedure described in detail by Lagos and
Paredes~\cite{LagosParedes}. It should be mentioned that it is
obtained using some rather crude hypotheses. However, the results
clearly exhibit what is most relevant: that the observed spectral
features can be explained by the excitation of three sharply defined
energy levels.  The theoretical curve of Fig.~\ref{Fig3} assumes that
the absorbing orbital is in a tetrahedral symmetry environment, and
that the vibrational modes are the bulk substrate modes. More accurate
results would demand a more detailed study, in particular the
incorporation of surface modes, but the present estimate is sufficient
for the purposes of what we intend to convey. Our point is that the
absorption curves can be explained by a finite number of sharp energy
levels, as opposed to a finite width energy band. Moreover, the
absorption spectra of the semiconducting CPs are quite similar. The
largest wavelength peak always has a shoulder, and in general the
spectra exhibit two main peaks clearly attributable to the two flat
bands, as derived above.  To obtain precise values for the discrete
energy levels above the HOMO energy from the UV--Vis excitation
spectra demands the deconvolution of the vibrational modes which widen
the spectral lines~\cite{LagosParedes}.

\begin{figure}[h!]
\begin{center}
\includegraphics[width=85mm]{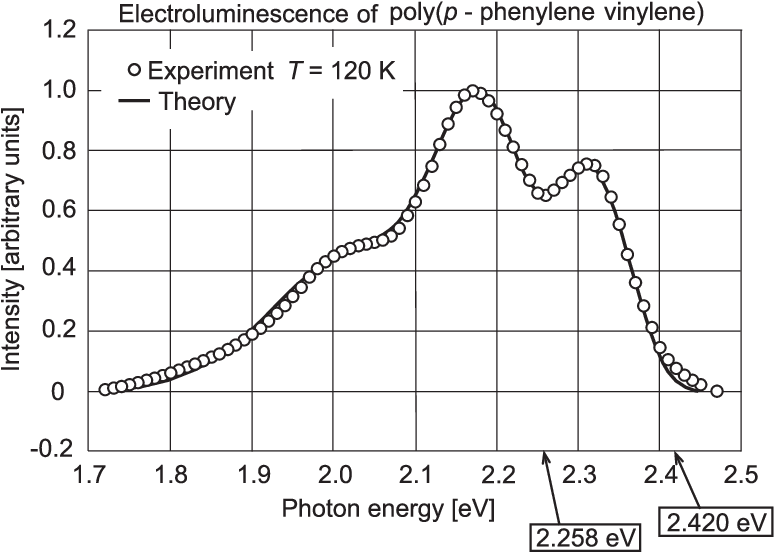}
\caption{\label{Fig4} Electroluminescent emission by an organic diode
made of the conjugate polymer poly(p--phenylene vinylene). Circles
represent the classical experimental work of Burroughes et
al.~\cite{Burroughes} and the solid line a calculation considering the
emission by two electronic energy levels of 2.258 and 2.420 eV. The
peaks are shifted, widened and split by multi--phonon events.}
\end{center}
\end{figure}

Fig.~\ref{Fig4} shows the emission spectrum of an electroluminiscent
device made of the CP poly(p--phenylene vinylene), published in a
seminal paper of Burroughes et al.~\cite{Burroughes}.  Open circles
describe the experimental results at a temperature $T=120\,\text{K}$,
and the solid line is the emission spectrum calculated with the same
procedure described by Lagos and Paredes~\cite{LagosParedes}, in
relation to Fig.~3. Phonon widening and shift of the spectral maxima
are particularly strong. The spectrum shows three peaks, but the
calculation assumes only two electronic energy levels, corresponding
to 2.258 and 2.420 eV, shown in Fig.~\ref{Fig4} with an arrow pointing
to their positions along the energy axis. The large Stokes shifts
caused by the vibration modes are quite evident, and is the cause of the
split of the spectral feature split into two maxima.

\section{Summary and conclusions}
\label{sec:concl}

We developed a model for covalent conjugate polymers including the
interaction between $\pi$--bonds. It is based on two hypothesis: i)~that
the covalent character of the $\pi$--bonds, forming stable octets,
is relevant in transport processes. This may be seen as a sort of pairing in
coordinate space. ii)~The explicit introduction of the repulsion between
the $\pi$--bonds of nearest neighbor sites.

The localized character of the two--particle states is not
inconsistent with their capability for transporting momentum, energy
and charge. We formulate a many--body model leading to transport under
these conditions.  A few very narrow energy bands with no dispersion
are predicted in the gap between the energies of the HOMO and the
LUMO. The novel bands are able to host many electron pairs that obey
Bose--Einstein statistics.  The potential for charge transport is very
large. In fact, charge transport in inorganic conductors is due to
electrons in a narrow fringe around the Fermi energy, and it turns out
that in this model the overwhelming  majority of the $\pi$--orbitals does
participate in the conduction.

Due to the localized nature of bonding in non-conjugated polymers, the
interactions between a monomer's constituent parts lead to a discrete
and well-defined spectrum of energy levels.  These sharp energy levels
become finite width energy bands when the monomers bond together to
form a polymer of considerable length. Therefore, in principle, the
observation of spectral lines of well defined energy, that originate
in sharp energy levels of the spectra of non--conjugated polymers, are
not expected. Traditionally, luminescence from polymeric luminogens is
attributed to the presence of conjugated structures~\cite{Lai}. In
this paper it is shown that the bands of conjugated polymers are due
to transitions between discrete energy levels of sharply defined
energy. The line width and shape observed in the optical spectra are
determined by the strong coupling with the vibrational modes,
particularly the acoustic ones. These discrete energy levels are due
to $\pi$--orbital repulsion. The competition of the next nearest
neighbouring $\pi$--orbitals that try to occupy the intermediate
almost empty site, and the strong repulsion between them, generates an
effect similar to the energy and momentum transmission in a hard
elastic ball line~\cite{Lehman}. Therefore, charge conduction and the
observation of absorption and emission spectral features are evidence
for the existence of a few sharp energy levels, in the energy gap
above the HOMO energy, and are closely related features of conjugated
polymers.

\vskip 1cm
\noindent {\bf Acknowledgements} MK was partially supported by
Financiamiento Basal para Centro Investigaci\'on Avanzada ANID, grant
CIA25002.

\end{document}